\documentclass[12pt]{article}
\usepackage{epsfig}\parskip 5pt plus 1pt
\usepackage{bbm}
\usepackage{amsmath}
\usepackage{amssymb}
\usepackage{amsfonts}
\usepackage{mathrsfs}
\usepackage{graphicx}
\newcommand{\nn}{\nonumber}

\newcommand{\dv}{\partial\hspace{-7pt}\slash}
\newcommand{\Dv}{D\hspace{-7pt}\slash}
\newcommand{\<}{\langle}
\renewcommand{\>}{\rangle}
\newcommand{\be}{\begin{equation}}
\newcommand{\ee}{\end{equation}}
\newcommand{\bea}{\begin{eqnarray}}
\newcommand{\eea}{\end{eqnarray}}
\begin{document}
\begin{titlepage}
\vspace*{-1cm}
%\phantom{hep-ph/0607020} 
%\hfill{LOCAL-PREPRINT-NUMBER}
\flushright{FTUAM-06-8 \\ IFT-UAM/CSIC-06-30}
\vskip 2.0cm
\begin{center}
{\Large\bf
Unitarity of the Leptonic Mixing Matrix
}
\end{center}
\vskip 1.0  cm
\begin{center}
{\large S. Antusch}~\footnote{antusch@delta.ft.uam.es},
{\large C. Biggio}~\footnote{carla.biggio@uam.es},
{\large E. Fern\'andez-Mart\'\i nez}~\footnote{enrique.martinez@uam.es},\\
{\large M.B. Gavela}~\footnote{gavela@delta.ft.uam.es} and 
{\large J. L\'opez-Pav\'on}~\footnote{jacobo.lopez@uam.es}
\\
\vskip .1cm
Departamento de F\'\i sica Te\'orica and Instituto de F\'\i sica Te\'orica,
\\
Universidad Aut\'onoma de Madrid, 28049 Cantoblanco, Madrid, Spain
\end{center}
\vskip 0.5cm
\begin{abstract}
\noindent

We determine the elements of the leptonic mixing matrix, without
assuming unitarity, combining data from neutrino oscillation
experiments and weak decays. To that end, we first develop a formalism
for studying neutrino oscillations in vacuum and matter when the
leptonic mixing matrix is not unitary.  To be conservative, only three
light neutrino species are considered, whose propagation is
generically affected by non-unitary effects.  Precision improvements
within future facilities are discussed as well.

\end{abstract}
\end{titlepage}
\setcounter{footnote}{0}
\vskip2truecm
\newpage
\section{Introduction}
\label{intro}

The experimental observation of neutrino oscillations both in the solar  \cite{exp_solares}-\cite{CHOOZ}
and atmospheric \cite{exp_atmo}-\cite{Minos}  
regimes constitutes evidence for new physics beyond the Standard Model of particle physics (SM).  Its favorite interpretation is neutrino masses, which allow flavour mixing. Neutrino masses and mixings have become an essential piece in the understanding of flavour dynamics.  The determination of the absolute scale of neutrino masses, of its character -whether Dirac or Majorana-, as well as a deep and precise understanding of leptonic mixing and CP violation, are all complementary milestones required to unravel the flavour puzzle, which is one of the most important goals of elementary particle physics.

Neutrinos are massless in the SM, since there are no right-handed neutrinos and the theory accidentally conserves $(B-L)$, where $B$ and $L$ denote baryon and lepton number, respectively. Neutrino masses can be Dirac or Majorana. Pure Dirac neutrino masses added to the SM are disfavored theoretically for the following reasons: 1) They worsen the problem of the large mass hierarchy between neutrinos and other fermions, with Yukawa couplings down to $11$ orders of magnitude smaller for the former. 2) They are unnatural in the 't Hooft sense, as Majorana mass terms of right-handed neutrinos are not forbidden by the SM gauge symmetry; Dirac neutrinos require to impose the {\it ad hoc} assumption  that the Lagrangian conserves $L$, and consequently $(B-L)$ so long as $B$ is conserved. 3) Such conservation precludes
the simplest scenarios for the generation of the baryon asymmetry of the universe through leptogenesis~\cite{fy}, at energies above the electroweak scale.

 Non-zero neutrino masses are thus a signal of physics beyond the SM.  The  new scale associated to it, $M$, is likely to be much higher than the electroweak scale. When $M\,\gg\,M_Z$, the effects of new physics 
 at  the high energy scale can be parameterized at low energies, without loss of generality, by  an effective Lagrangian including: 
 \begin{itemize}
 \item Corrections to the parameters of the SM Lagrangian.
 \item The addition to the SM Lagrangian of a tower of {\it non-renormalizable} -higher dimensional- effective operators,  which are $SU(3) \times SU(2) \times U(1)$ gauge invariant. Their coefficients  are suppressed by inverse powers of the large energy scale $M$.
\end{itemize}
The relative intensity of the different higher-dimensional effective operators is model 
 dependent. The SM couplings themselves are expected to be generically modified, though, in any scenario involving new physics at high energy. We concentrate on modifications to those couplings, in the present work.
 
   Among the SM couplings, the PMNS mixing matrix is introduced in connection with charged current interactions of leptons. Its departure from the unit matrix is the origin of all leptonic mixing and putative CP-violation effects in neutrino physics, which are essential building blocks of the flavour puzzle. In typical analyses of experimental data, the mixing matrix is assumed to be unitary.
   
 The complete Lagrangian including new physics is necessarily unitary and probability is conserved. 
 Unitarity violation is  one typical low-energy signal of models of new physics, though,  when data are analyzed in the framework of the SM gauge group  with three light species of quarks, leptons and neutrinos, plus mass terms for the latter. Notice that for the analogous matrix in the quark sector, the CKM matrix, deviations from unitarity  are considered a good window for physics beyond the SM and extensively studied.

    An evident possible source of non-unitary effects in leptonic mixing is the hypothetical existence of more than three light neutrino species,   that is, light sterile~\cite{sterile} neutrinos. Their theoretical implementation typically requires strong fine-tunings, though, and no definite signal exists for their existence 
\cite{LSND}-\cite{Stancu:2006gv}. To be conservative, we will disregard such possibility in the analysis below. 
 Would light sterile neutrinos turn out to  be indeed present  in nature and detected, its presence in mixing processes would obviously lead to stronger signals than those developed in this paper.

Even in the framework of only three light neutrinos, extensions of the SM postulated in order to generate the observed small neutrino masses typically produce a leptonic mixing matrix which is non-unitary. 
For instance, any theory in which the neutrino mass matrix turns out to 
be part of a larger matrix,  which involves heavy fields, may  generically result in an effective low-energy non-unitary mixing matrix for the light leptonic fields~\cite{Zralek}-\cite{newph}.
A paradigmatic example is the {\it seesaw mechanism}~\cite{seesaw}.
In it, light neutrino masses take natural (i.e. non-fine-tuned) values, in contrast to all other fermion masses except the top quark mass. Heavy sterile (right-handed) neutrinos, with masses far above the electroweak scale, are introduced at the new high energy scale. Although the non-unitary mixing induced at low energies in the canonical  -type I- seesaw model is typically expected to be too small for detection, this is not necessarily true for variants of the seesaw mechanism, or other theories beyond the SM.

At the same time, we are about to enter an era of high precision neutrino physics. With on-going and forthcoming experiments, as well as the future facilities under discussion,  future neutrino oscillation experiments will be aiming at a measurement of the last unknown leptonic mixing angle $\theta_{13}$, even if very small, as well as of leptonic CP violation. It is pertinent to ask whether such precision can shed further light on unitarity. Departures from it would probe the new physics behind.

 We will thus relax here the assumption of unitarity in the low-energy leptonic mixing matrix of weak interactions, and let data rather freely tell us up to what point the measured elements of the mixing matrix  arrange themselves in a unitary pattern. This will allow to identify the 
  less constrained windows in flavour space and thus the most sensitive ones, as regards  new physics. 
  
  We will not work in any concrete model of neutrino masses. Nevertheless, as only neutrino masses clearly signal  new physics -in contrast to masses for charged leptons or the rest of the fermions-,
it is plausible that the new physics behind sneaks through at low energies primarily through its effects on neutrino propagation. That is,  
 in this first work we implicitly assume that  
 the physics of fields other than neutrinos will be that of the SM.
 We can summarize then our approximations  on a set-up  that we will
  dub { \it \bf  Minimal Unitarity Violation} (${\bf MUV}$), based on the following assumptions: 
   \begin{itemize}
   \item  Sources of non-unitarity are allowed in those terms in the SM Lagrangian which involve neutrinos.
   \item Only three light neutrino species are considered.
   \end{itemize}
  Leptonic and semileptonic decays, together with neutrino oscillations, will be analyzed in this minimal set-up. Supplementary non-unitary contributions to physical transitions can 
  result from new physics affecting the SM couplings, for fields other than neutrinos and/or higher-dimensional operators in the effective Lagrangian.
  Barring extreme fine-tunings, they should not affect the order of magnitude of the results obtained in the MUV scheme, though.

The paper is organized as follows:
 Sec.\ \ref{N}  defines the framework and introduces the non-unitary mixing matrix $N$ which replaces the unitary PMNS matrix. In Sec.\ \ref{oscillations}, a formalism is developed for the study of  neutrino oscillations in vacuum and matter, with non-unitary leptonic mixing.
Secs.\ \ref{fits}  and \ \ref{decays}  deal with data re-analyzed in the MUV scheme: 
present data from neutrino oscillation experiments are considered in Sec.\ \ref{fits} and the mixing matrix resulting from their analysis is obtained, while the unitarity constraints resulting from $W$ and $Z$-decay data, lepton universality tests and rare charged lepton decays are presented in  Sec.\ \ref{decays}.  The final mixing matrix resulting from the combination of oscillation and weak decays data is presented in Sec.\ \ref{totalmixmatrix}. In Sec.\ \ref{future}, the impact of future experiments on the results obtained in the previous sections is studied. In
Sec.\ \ref{conclu} we conclude.

%%%%%%   section 2   %%%%%%%%%%%%%%%%%%%%%%%%%%%%%%%%%%%%%%%%%%%%%%%
\section{Non-unitary mixing matrix}
\label{N}

Our aim is to substitute the usual leptonic unitary matrix $U_{PMNS}$ by a non-unitary one. 
To be definite, we will analyze the consequences of an effective low-energy Lagrangian which, in the mass basis, reads

 \bea
\label{eff-lagr}
{\cal L}^{eff}&=&
\frac{1}{2}\,\left(
\bar{\nu}_ii\,{\partial\hspace{-6pt}\slash}\,\nu_i
-\,
\overline{{\nu}^c}_im_{i}\,\nu_i  +\,h.c.\right)\,
-\,\frac{g}{2\sqrt{2}}\,
 ( W^+_\mu\,\bar{l}_\alpha\,\gamma_\mu\,(1-\gamma_5)\,N_{\alpha i}\,\nu_i
   \,+\,h.c.) \nn\\
&&-\,\frac{g}{2 \cos\theta_W}\,
 ( Z_\mu\,\bar{\nu}_i\,\gamma^\mu\,(1-\gamma_5)\,
     (N^{\dagger}N)_{ij}\,\nu_j\,+\,h.c.)\,
+\,\dots \, 
\eea
 where $\nu_i$ denotes four-component left-handed fields. 
 Eq.~(\ref{eff-lagr}) is the usual Lagrangian for neutrinos in the mass basis, albeit with the unitary matrix $U_{PMNS}$
 in the charged current  substituted by a general
non-unitary matrix $N$. The neutral current coupling has been modified as well, as expected in general from the non-unitarity of $N$. Notice that the Lagrangian includes  a Majorana mass term for neutrinos, for the sake of definiteness,
  although for our numerical analysis below it would make no difference to consider neutrinos of the Dirac type.

It is necessary to clarify the relation between mass and flavour eigenstates. With non-unitarity present, both the mass and flavour bases cannot be orthonormal.
Indeed, 
 $N$ connects the quantum fields in the mass basis with those in the flavour basis\footnote{Throughout the paper, Greek (Latin) indices label the flavour (mass) basis.} where the weak couplings are diagonal,
\be
\label{field-transf1}
\nu_{\alpha} = N_{\alpha i}\, \nu_{i}  \, .
\ee
While the canonical kinetic terms in Eq.~(\ref{eff-lagr})  give rise to orthonormal mass eigenstates,
\be
\label{ortoij}
\langle\nu_i | \nu_j\rangle=\delta_{ij} \, ,
\ee
 consistency between quantum states and fields requires the identification~\cite{Giunti}
\be
\label{states-rel}
|\nu_{\alpha} \rangle = \frac{1}{\sqrt{(NN^\dagger)_{\alpha\alpha}}}
\sum_i\,N_{\alpha i}^*\,|\nu_i\rangle
\equiv\sum_i \tilde{N}_{\alpha i}^* |\nu_i\rangle \, ,
\ee
where, on the right-hand side, the normalization factor has been absorbed in the definition of $\tilde{N}$. This equation is 
to be compared with Eq.~(\ref{field-transf1}). It follows from it that flavour eigenstates are no more orthogonal\footnote{Notice that these are effective low-energy flavour eigenstates. In the corresponding complete -high energy- hypothetical theory, it should be possible to define an orthonormal flavour basis.}:
\be
\label{noortoflav}
\langle\nu_\beta | \nu_\alpha\rangle 
= (\tilde{N}\tilde{N}^\dagger)_{\beta\alpha} \neq \delta_{\alpha\beta} \, ,
\ee
which will induce relevant physical effects, as it will be shown later on.

Prior to any predictions for physical transitions,  the Lagrangian must be renormalized, determining from experiment its free parameters. In particular, the weak coupling  in 
Eq.~(\ref{eff-lagr}) differs from the SM expression. Accordingly, the Fermi constant measured in experiments cannot be identified anymore with the SM tree level combination $G_F = \sqrt{2} g^2 / (8 M_W^2)$, due to non-unitarity. For instance, the Fermi constant $G^M_F$ extracted from the decay $\mu \rightarrow \nu_\mu e \bar{\nu}_e$ is related to $G_F$ by
\be
\label{Eq:GFM}
G_F = 
\frac{G_F^M}
{ \sqrt{(NN^\dagger)_{ee} (NN^\dagger)_{\mu\mu} }}\, .
\ee
The rest of the parameters of the Lagrangian coincide with those in the standard treatment.

%%%%%%%%%%%%
\subsubsection*{The effective Lagrangian in the flavour basis}

Let us write the Lagrangian in Eq.~(\ref{eff-lagr}) in the flavour basis. Upon the transformation in Eq.~(\ref{field-transf1}), it follows that
\bea
\label{gen-lagr}
{\cal L}^{eff}&=&\frac{1}{2}
\left(i\,\bar{\nu}_{\alpha}\,\,{\partial\hspace{-6pt}\slash}\,
(NN^\dagger)^{-1}_{\alpha\beta}
 \,\,\nu_{\beta}\,
- \,\overline{{\nu}^c}_{\alpha}\,
[(N^{-1})^t mN^{-1}]_{\alpha\beta}
\,\nu_{\beta}\,+\,h.c.\right) \nn\\
& & -\,\frac{g}{2\sqrt{2}}\,
 \big( W^+_\mu\,\bar{l}_\alpha\,\gamma^\mu\,(1-\gamma_5)\,\nu_\alpha
       +\,h.c.\big)\nn\\
& &-\, \frac{g}{2 \cos\theta_W}\, 
 \big( Z_\mu\,\bar{\nu}_\alpha\,\gamma^\mu\,(1-\gamma_5)\,\nu_\alpha
       +\,h.c.\big)\,
+\, \dots \, ,
\eea
where $m\equiv \mbox{diag}(m_1,m_2,m_3)$.
In this basis, weak couplings are diagonal and a general neutrino mass matrix and kinetic term have appeared. A neutrino mass term  can always be diagonalised by a unitary transformation (or bi-unitary in the case of Dirac neutrinos), leading to a unitary contribution to the mixing matrix, as in the standard treatment. Kinetic terms must be both diagonalized {\it and normalized}, though, so as to obtain canonical kinetic energies. 

The hypothetical different normalizations induced in the kinetic energy of neutrino fields by the new physics, is the key point to obtain non-unitary effects in neutrino mixing, in the MUV scheme. Whenever at least two normalizations of neutrino fields differ, a non-unitary weak mixing matrix follows.

For the sake of physics intuition, let us discuss the theoretical consistency and implications behind the Lagrangian 
in Eqs.~(\ref{eff-lagr}) and (\ref{gen-lagr}), although the hypothetical reader mainly interested in the numerical analysis is invited to proceed directly to the next Section.

 A pertinent question is whether  there exists a $SU(2)\times U(1)$ invariant formulation of the 
 MUV formulation above, as it should.
Consider a generic effective  Lagrangian valid at energies less than a high scale $M$ of new physics, $M\gg M_Z$, resulting after integrating out the heavy fields present above such scale. The effective Lagrangian has a power series 
expansion in $1/M$ of the form
\bea
\label{Lag}
{\cal L}^{eff}={\cal L}_{SM}\,+\,{\delta \cal L}^{d=5}\,+\,{\delta \cal L}^{d=6}+\, \dots \, ,
\eea
where ${\cal L}_{SM}$ contains all $SU(3)\times SU(2) \times U(1)$ invariant operators of dimension $d\le4$ and the gauge invariant operators  of $d>4$, constructed from the SM fields, account for the physics effects of the heavy fields at energies $< M$.
After electroweak spontaneous symmetry breaking (SSB), the operators with 
 $d>4$ will give corrections to the couplings present in the SM Lagrangian {\it and} also produce new exotic couplings. 
 
  The question is whether there exist  $SU(2)\times U(1)$ gauge invariant formulations of the Lagrangian 
   which give rise, after SSB, to  flavour mixing corrections  to the  neutrino kinetic energy, as discussed here. 
  This is the case indeed. There exists for instance a $d=6$ gauge invariant operator which precisely results in the corrections in the MUV scheme.  It is the operator characteristic of 
 the canonical seesaw model\footnote{ In addition to the well-known $d=5$ operator responsible for neutrino masses, $\delta{\cal L}^{d=5} = \frac{1}{2}\, c_{\alpha \beta}^{d=5} \,
\left( \overline{L^c}_{\alpha} \tilde \phi^* \right) \left(
\tilde \phi^\dagger \, L_{ \beta} \right) +  h.c.\,$, where $c_{\alpha \beta}^{d=5}$ is the coefficient matrix of $\cal{O}$$(1/M)$.}   and its generalizations~\cite{Broncano:2002rw}, and also of some extra-dimensional constructions~\cite{extradim},
 \be\label{d6}
\delta{\cal L}^{d=6} = c^{d=6}_{\alpha \beta} \, \left( \overline{L}_{\alpha} \tilde \phi
\right) i \dv \left( \tilde \phi^\dagger L_{ \beta} \right),
\ee 
where $L$ denotes left-handed leptonic doublets\footnote{ As the factors in parenthesis in this equation are singlets of $SU(2)\times U(1)$, ${\partial\hspace{-6pt}\slash}$ is tantamount to $\Dv$ in this operator.}, $c^{d=6}$ is the -model dependent- coefficient matrix of $\cal{O}$$(1/M^2)$ and $\tilde \phi$ is related to
the standard Higgs doublet $\phi$ by $\tilde \phi = i \tau_{2} \phi^*$.

 Other $SU(2)\times U(1)$ invariant operators may be written as well,
  resulting generically -after SSB- in corrections to both the neutrino and charged leptons kinetic energy. There exist even $SU(2)\times U(1)$ invariant operators contributing to the charged lepton kinetic energy and not to that of neutrinos, such as
 \be\label{d6jacobo}
\delta{\cal L}^{d=6} = c^{\prime \, d=6}_{\alpha \beta} \, \left( \overline{L}_{\alpha}  \phi
\right) i \Dv \left(  \phi^\dagger L_{ \beta} \right),
\ee  
with $c^{\prime\, d=6}$ being the coefficient matrix of order $\cal{O}$$(1/M^2)$. 
Theories with  Yukawa couplings to heavy extra fermions -be it of Dirac or Majorana type- can easily give rise to such effective couplings at low energy.
   Such an operator leads -after SSB- to a Lagrangian  with the same couplings to the $W$ boson as in Eq.~(\ref{eff-lagr}), albeit with corrections in the Z-charged lepton couplings instead of in the Z-neutrino ones.  It means that all results obtained below from $W$ exchange alone would also hold for the purpose of constraining such theories.
   
   Many other gauge invariant operators~\cite{buch} can contribute to the MUV physics discussed here. The last comments above are only a digression, to illustrate the
  rather general application realm of our treatment, although we will keep our focus on neutrinos as stated.

   In summary, a generic model is expected to give  rise after SSB to modifications of the standard couplings, as well as new exotic ones.
   In the minimal scheme analyzed in this paper, MUV, only the former will be taken into account  and more precisely only those couplings involving neutrinos as specified above, as
 it is natural to expect that the latter will be specially sensitive to the new physics responsible for neutrino masses. This simplification should provide a sensible estimation of the best windows for non-unitarity, unless extreme fine-tunings and cancellations occur between the different  type of contributions, in some hypothetical model.
 
%%%%%%   section 3   %%%%%%%%%%%%%%%%%%%%%%%%%%%%%%%%%%%%%%%%%%%%%%%

\section{Neutrino oscillations without unitarity}
\label{oscillations}

Let us consider now the impact of the Lagrangian in Eq.~(\ref{eff-lagr}) on neutrino oscillations, both in vacuum and in matter.

\subsection{Vacuum oscillations}
\label{vacuum}

Consider free neutrino propagation, described by the free Hamiltonian $\hat{H}^{free}$,  resulting from the first two terms in the Lagrangian, Eq.~(\ref{eff-lagr}).
The time evolution of mass eigenstates follows the usual pattern. Indeed,
\bea
\label{evi}
i \frac{d}{dt} |\nu_i \rangle = \hat{H}^{free}\, |\nu_i \rangle\,
\eea
and because of the orthogonality of the mass basis, 
\bea
 \langle \nu_j | \hat{H}^{free} | \nu_i \rangle\equiv\delta_{ij}\,E_i\,,
\label{Hab}
\eea
where $E_i$ are the eigenvalues. Using now the completeness relation,
$\sum_j |\nu_j \rangle \langle \nu_j |=1$, Eq.~(\ref{evi}) reads:
\be
\label{evmass}
i \frac{d}{dt} |\nu_i \rangle = \sum_j |\nu_j \rangle \langle \nu_j | \hat{H}^{free} |\nu_i \rangle = E_i|\nu_i \rangle\, ,
\ee
which is the usual time propagation for free states.

Consider now instead the free evolution in the flavour basis, which is {\it not} orthonormal 
and  for which there is not the usual completeness relation, as $\sum_\alpha |\nu_\alpha \rangle \langle \nu_\alpha |\ne1$. The time evolution is given by 
\bea
\label{eval}
i \frac{d}{dt} |\nu_\alpha \rangle = \hat{H}^{free}\, |\nu_\alpha \rangle\, ,
\eea
which, using the orthogonality and completeness of the mass basis, results into 
\begin{eqnarray}
\label{evflav}
i \frac{d}{dt} |\nu_{\alpha} \rangle 
= 
\sum_j |\nu_j\rangle \langle \nu_j|\hat{H}^{free}|\nu_{\alpha}\rangle
=
\sum_\beta (\tilde{N}^*\,E\, (\tilde{N^{*}})^{-1} )_{\alpha \beta}|\nu_{\beta}\rangle\,,
\end{eqnarray}
where $E\equiv \mbox{diag}(E_1,E_2,E_3)$.
This is to be compared with the
$N$-dependence of the matrix elements between flavour eigenstates, given by
 \bea
 \label{matrix_elem_fl}
  \langle \nu_\beta | \hat{H}^{free} | \nu_\alpha \rangle =
   \,(\tilde{N}^*\,E\,\tilde{N}^t)_{\alpha \beta}\,.
\eea
That is,  the evolution in flavour space is driven by the combination $(\,\tilde{N}^*\,E\, (\tilde{N^{*}})^{-1}\, )$ and {\bf not} by the product
$(\tilde{N}^*\,E\,\tilde{N}^t)$ appearing in Eq.~(\ref{matrix_elem_fl}), in contrast to  the customary  expression in standard -unitary- treatments. Because of the non-unitarity of $N$ both expressions are no more equivalent.
Technically, this is a key point in the different results for the non-standard case, to be obtained below.

 Notice, moreover, that the combination  $(\,\tilde{N}^*\,E\, (\tilde{N^{*}})^{-1}\, )$  is not  Hermitian, 
even if the free Hamiltonian itself is Hermitian. This in turn implies that the
evolution of flavour bra states, $\langle \nu_{\alpha} |$, differs from the
evolution of the flavour kets, leading both to the same probability equation, as they should.

 The analysis of free propagation in space is analogous to that for time evolution described above and we will not repeat it in detail.
 Flavour eigenstates, after a distance $L$, transform into  
\be
\label{nua(L)}
|\nu_{\alpha}(L)\>\,=\,\sum_{i\gamma}\,\tilde{N}^*_{\alpha i}\,e^{i\,P_i\,L}\,
                      (\tilde{N}^*)^{-1}_{i\gamma}\, |\nu_{\gamma}\> \, ,
\ee
where $P_i$ are the momentum eigenvalues, $P_i=\sqrt{E_i^2-m_i^2}$. 
The oscillation probability after traveling a distance $L$ can now be obtained, 
\bea
\label{prob}
P_{\nu_\alpha \nu_\beta}(E,L)
&\equiv&|\<\nu_{\beta}|\nu_{\alpha}(L)\>|^2\,=\,
\frac{|\sum_i N^*_{\alpha i}\,e^{i\,P_i\,L}\,N_{\beta i}|^2}
{(NN^\dagger)_{\alpha\alpha}(NN^\dagger)_{\beta\beta}}=\\[0.2cm]
\label{explicitprob}
&=&\frac{1}
{(NN^\dagger)_{\alpha\alpha}(NN^\dagger)_{\beta\beta}}\bigg[
\sum_i |N_{\alpha i}|^2|N_{\beta i}|^2+\nn\\[0.1cm]
&&+2 Re\left\{N_{\alpha 1}N^*_{\alpha 2}
      N^*_{\beta 1}N_{\beta 2}\right\} \cos \Delta_{12}
+2 Im\left\{N_{\alpha 1}N^*_{\alpha 2}
      N^*_{\beta 1}N_{\beta 2}\right\} \sin \Delta_{12} +\nn\\[0.2cm]
&&+\,2 Re\left\{N_{\alpha 2}N^*_{\alpha 3}
      N^*_{\beta 2}N_{\beta 3}\right\} \cos \Delta_{23}
+2 Im\left\{N_{\alpha 2}N^*_{\alpha 3}
      N^*_{\beta 2}N_{\beta 3}\right\} \sin \Delta_{23} +\nn\\
&&+\,2 Re\left\{N_{\alpha 3}N^*_{\alpha 1}
      N^*_{\beta 3}N_{\beta 1}\right\} \cos \Delta_{31}
+2 Im\left\{N_{\alpha 3}N^*_{\alpha 1}
      N^*_{\beta 3}N_{\beta 1}\right\} \sin \Delta_{31} \bigg]\,,\nn
\eea
where $\Delta_{ij}=\Delta m^2_{ij}L/2E$, with $\Delta
m^2_{ij}=m^2_i-m^2_j$, as usual. Written in this way, the expression is easily seen to reduce to the standard one if $N$ was unitary, as it should.

 The first very important consequence of Eq.~(\ref{prob}) is that  the non-unitarity of $N$
is shown to generate a ``zero-distance'' effect \cite{zerodistance}, i.e.\ a flavour transition already at the source before oscillations can take place. Indeed, for $L=0$, it follows that
\be
\label{zerodist}
P_{\nu_\alpha \nu_\beta}(E, L=0) = \frac{|(NN^\dagger)_{\beta\alpha}|^2}
{ (NN^\dagger)_{\beta\beta}\,(NN^\dagger)_{\alpha\alpha}   }\neq 0 \, ,
\ee
an effect that can be tested in near detectors, thus setting strong limits on unitarity as we will see later.
Nevertheless, due to non-unitarity, the probability as defined in Eq.~(\ref{prob}) does not sum up to a total probability of 100\%.
 To make contact with data, let us  discuss the implications of our treatment  for the production and detection cross sections and, finally, for the number of events detected in a given experiment.

\subsubsection{Production/detection cross sections and widths}

 The non-unitarity of the mixing matrix $N$ implies the following corrections, for processes computed at tree-level:
  
 \begin{itemize} 
 
 \item Charged current (CC) cross-sections and fluxes involving only one neutrino flavour $\alpha$ are given by
 
 \be
\label{xcc1}
 \sigma^{CC}_\alpha\, = \sigma_\alpha^{CC(SM)}\,(N N^\dagger)_{\alpha\alpha}\,, \hspace{1cm} \, \frac{d\Phi_\alpha^{CC}}{dE} = \frac{d\Phi_\alpha^{CC(SM)}}{dE}\,
(NN^\dagger)_{\alpha\alpha}   \; , 
\ee
where $\sigma^{CC(SM)}_\alpha$ and $\Phi_\alpha^{CC(SM)}$ are the SM cross section and flux, respectively. The same correction factor affects decay widths involving one  neutrino flavour.

\item Charged current cross sections involving two neutrino flavours, $\alpha, \beta$, will be modified into
\be
\label{xcc2}
 \sigma^{CC}_{\alpha,\beta} = \sigma^{CC(SM)}_{\alpha,\beta}\,(N N^\dagger)_{\alpha\alpha}  \, (N N^\dagger)_{\beta\beta}  \, ,
\ee
with the same weight factor affecting widths or fluxes involving two neutrino flavours.
\item Neutral current (NC) processes are weighted by a different combination.
A decay width involving two neutrino mass eigenstates,
 $\nu_i,\nu_j$, is given by 
 \begin{eqnarray}
 \label{Zwidth}
\Gamma (Z \rightarrow \bar{\nu}_i \nu_j ) = \Gamma^{SM} (Z \rightarrow \bar{\nu}_i \nu_i ) \,|\,(N^\dagger N)_{ij}\,|^2\,.
\end{eqnarray}
Analogously, when detecting a neutrino $\nu_i$ through neutral current interactions, as in SNO,  modified cross sections will have to be considered,
\be
\label{xnc}
\sigma^{NC}_{i} = \sum_{j}\sigma^{NC(SM)}\,|\,(N^\dagger N)_{ij}\,|^2\,,
\ee
where the sum over $j$ is due to the fact that the final neutrino $\nu_j$ remains  undetected.
\end{itemize}

 \subsubsection{ Number of events}

The number of events in a detector located at a distance $L$ away from the source would be given, apart from backgrounds, by the convolution of the production flux, the oscillation probability, the detection cross-section and the detector efficiency, integrated over energy. In short,
\be
\label{eventos}
n_{events}\sim\int dE\,\frac{d\Phi_\alpha(E)}{dE}\,
P_{\nu_\alpha \nu_\beta}(E,L)\,\sigma_\beta(E)\,\epsilon(E) ,
\ee
where $d\Phi_\alpha(E)/dE$ is the neutrino flux, $\sigma_\beta(E)$ is the detection cross
section and $\epsilon(E)$ the detection efficiency.
In the presence of MUV, all factors in Eq.~(\ref{eventos})  should be corrected, as discussed above. It is easy to see that there are cancellations between the different $N$ dependent factors they exhibit.

For instance, for experiments in which both production and detection take place via charged currents, involving each one neutrino flavour, the denominator of $P_{\nu_\alpha \nu_\beta}$ -Eq.~(\ref{prob})- cancels the
correction factors in the flux and cross section, Eq.~(\ref{xcc1}). This allows
to express in this case the number of events simply as 
\bea
\label{eventos2}
n_{events}
\sim\int dE\,\frac{d\Phi^{CC(SM)}_\alpha(E)}{dE}
\hat{P}_{\nu_\alpha \nu_\beta}(L,E)\,
\sigma^{CC(SM)}_\beta(E)\,\epsilon(E) \,,
\eea
where
$\hat{P}_{\nu_\alpha \nu_\beta}(L,E)$ is the probability in Eq.~(\ref{prob}), amputated from the normalization factors in its denominator,
\bea
\hat{P}_{\nu_\alpha \nu_\beta}(L,E)
&\equiv&
|\sum_i N^*_{\alpha i}\,e^{i\,P_i\,L}\,N_{\beta i}|^2\,.
\label{correctedprob}
\eea
It turns out that, in practice, most  experiments extract the probabilities from the measured number of events, parameterized -via Monte Carlo simulations- in terms of the SM fluxes and cross sections, precisely as in Eq.~(\ref{eventos2}). Within MUV, their analysis thus provides a direct estimation of  $\hat{P}_{\nu_\alpha \nu_\beta}(L,E)$ in Eq.~(\ref{correctedprob}). This is the case for the very large number of experiments in which neutrinos are detected via charged current interactions and produced from decays of hadrons like $\pi$, $K$ or $\beta$ decays.

Obviously, there are exceptions.  For instance, if the neutrino flux
expected in the far detector of the previous example is not taken from
a Monte Carlo simulation, but from a direct measurement in a near
detector, the cancellation described in the previous paragraph would not be complete and extra $N$-dependent factors will have to be taken into account.

 Besides, when the production mechanism is not hadronic,
 but leptonic, as from $\mu$  or even $\tau$ decays, the fluxes would need two corrections instead of one, since the production involves two insertions of $N$, as in Eq.~(\ref{xcc2}). For instance,  the neutrino fluxes produced at a Neutrino Factory from $\mu$ decay should thus  be corrected by the factor $(NN^\dagger)_{\mu\mu}(NN^\dagger)_{ee}$.

 Finally,  the analysis of detection through neutral current processes is modified as well. Since such processes are sensitive to the sum of  all neutrino species, the number of events is given by
\bea
\label{eventos2amp}
n_{events}
\sim \int dE\,\frac{d\Phi^{CC(SM)}_\alpha(E)}{dE}
\sum_i \hat{P}_{\nu_\alpha \nu_i}(L,E)\,
\sigma^{NC}_i(E)\,\epsilon(E) \,,
\eea
with $\sigma^{NC}_i(E)$ as in Eq.~(\ref{xnc}) and -for instance for propagation in vacuum- 
\bea
\hat{P}_{\nu_\alpha \nu_i}(L,E)
&\equiv&
|N_{\alpha i}|^2\,.
\label{correctedprobi}
\eea

%%%%%%   subsection 3.2   %%%%%%%%%%%%%%%%%%%%%%

\subsection{Matter effects}
\label{matter}

When neutrinos pass through matter, they interact with the medium and the effect is a modification of the evolution and
consequently of the oscillation probability. Consider the usual interaction Lagrangian 
\be
\label{Lmatterfl}
-{\cal L}^{int}\,=\,\sqrt{2} G_F n_e \bar{\nu}_e \gamma_0 \nu_e -
\frac{1}{\sqrt{2}} G_F n_n \sum_\alpha \bar{\nu}_\alpha \gamma_0 \nu_\alpha \,,
\ee
where $n_e$ and $n_n$ are respectively the electron and neutron
densities. The first term corresponds to charged interactions,
while the second term corresponds to neutral interactions.
In the mass basis within MUV,  Eq.~(\ref{Lmatterfl}) reads 
\be
\label{Lmattermass}
-{\cal L}^{int}\,=\,
V_{CC} \sum_{i,j}N_{ei}^*N_{ej}\bar{\nu}_i \gamma_0 \nu_j -
V_{NC} \sum_{\alpha,i,j} N_{\alpha i}^*N_{\alpha j}\bar{\nu}_i \gamma_0\nu_j\,,
\ee
where $V_{CC}=\sqrt{2}\,G_F\,n_e$ and
$V_{NC}=\frac{1}{\sqrt{2}}\,G_F\,n_n$.  In order to estimate the time evolution of states passing through matter, consider the interaction Hamiltonian, $\hat{H}^{int}$, corresponding to Eqs.~(\ref{Lmatterfl}) and (\ref{Lmattermass}). Its matrix elements in the mass basis read
\be
\label{Hij-int}
H_{ij}^{int}\equiv
\langle\nu_j|\hat{H}^{int}|\nu_i \rangle
=V_{CC}N_{ei}N_{ej}^*-V_{NC}(N^{\dagger}N)_{ji}\, ,
\ee
or, in matrix notation 
\be
\label{H-int}
{H}^{int}\equiv \left[N^\dagger\,\mbox{diag}(V_{CC}-V_{NC},-V_{NC},-V_{NC})\,N\right]^t\, .
\ee
The evolution equation for mass eigenstates in matter is then given by
\be
\label{evmass-int}
i\,\frac{d}{dt} |\nu_{i}\rangle = \sum_j \left[
E\,+\,H^{int} \right]_{ij}  
 |\nu_j\rangle\, ,
\ee
where $E$ is the energy matrix for free eigenstates, introduced in Eq.~(\ref{evflav}).
In contrast, the evolution through matter of flavour eigenstates is given by
\be
\label{evflav-int}
i\,\frac{d}{dt} |\nu_{\alpha}\rangle = \sum_\beta \left[\tilde{N}^*
(E+H^{int})
(\tilde{N}^*)^{-1}\right]_{\alpha \beta}
 |\nu_{\beta}\rangle\,,
\ee
where again $(\tilde{N}^*)^{-1}$ cannot be traded by  $\tilde{N}^t$ -as it is usually done in the standard case- because  $N$ is not unitary.
  
  It is easy, although cumbersome, to write now explicitly the equations above for the three family case. To illustrate the main new effects, it is enough to write here explicitly  the effective flavour potential in the second term in Eq.~(\ref{evflav-int}), for the case of two
families:
\begin{eqnarray}
\label{Eflav-int}
\tilde{N}^*
H^{int} (\tilde{N}^*)^{-1}
&=&\tilde{N}^*\,N^t\, 
\left(\begin{array}{cc}
      V_{CC}-V_{NC} & 0\\
      0 & -V_{NC}
\end{array}\right)\,
N^*\,(\tilde{N}^*)^{-1}=\\[0.1cm]
&=&\left(\begin{array}{cc}
 (V_{CC}-V_{NC})(NN^\dagger)_{ee}
 & -V_{NC} \sqrt{\frac{(NN^\dagger)_{\mu\mu}}{(NN^\dagger)_{ee}}} 
    (NN^\dagger)_{ \mu e}\\
 (V_{CC}-V_{NC})\sqrt{\frac{(NN^\dagger)_{ee}}{(NN^\dagger)_{\mu\mu}}} 
    (NN^\dagger)_{e\mu}
 & -V_{NC}(NN^\dagger)_{\mu \mu}
\end{array}\right)\,.\nn
\end{eqnarray}
Consequently, MUV results generically in exotic couplings in the
evolution through matter. This effective potential is not diagonal, in
contrast to the unitary case.  Moreover, the neutral current
contribution can not be rewritten as a global phase in the evolution
equation and thus it contributes to the oscillation probabilities.

%%%%%%   section 5   %%%%%%%%%%%%%%%%%%%%%%%%%%%%%%%%%%%%%%%%%%%%%%%

%%%%%%%%%
\section{Matrix elements from neutrino oscillations}
\label{fits}

We will use now the most relevant data on neutrino oscillations, to
determine the elements of the mixing matrix, without assuming
unitarity. In this first work, we do not perform an exhaustive
analysis of {\it all} existing oscillation data; our aim is rather to
estimate what is the role played by the different experiments in
constraining the matrix elements.

All positive oscillation signals available nowadays\footnote{Except
for the LSND experiment \cite{LSND} which is currently being tested by
MiniBooNe \cite{Stancu:2006gv} and which we will not consider in this
paper.} correspond to disappearance experiments
\cite{exp_solares}-\cite{Minos}. Since the disappearance oscillation
probability in vacuum is given by
\bea
\nonumber
\hat{P}_{\nu_\alpha \nu_\alpha}
&=&
|N_{\alpha 1}|^4 + |N_{\alpha 2}|^4 + |N_{\alpha 3}|^4 +2|N_{\alpha1}|^2|N_{\alpha2}|^2 \cos \Delta_{12} \\
&+&2|N_{\alpha1}|^2|N_{\alpha3}|^2 \cos \Delta_{13}+2|N_{\alpha2}|^2|N_{\alpha3}|^2 \cos \Delta_{23}\, ,
\label{disap}
\eea
 disappearance experiments  may provide information on
the moduli of elements, while phases will remain unknown, as in the
unitary case. Furthermore, as no $\nu_\tau$ disappearance experiment
has been performed yet, this type of vacuum experiments will only
constrain the elements of the $e$ and the $\mu$-rows, as
Eq.~(\ref{disap}) indicates.  Nevertheless, the no-oscillation results
from some appearance experiments will also provide useful non-unitarity
constraints.

 \subsubsection*{Vacuum oscillations}
  
The exact appearance and disappearance probabilities in vacuum, Eqs.~(\ref{prob}) and (\ref{correctedprob}), will be used in the numerical analysis.
 To illustrate the discussion, the amputated probabilities can be approximated as follows, though, for some experiments studied below,  depending on the range of $L/E$:

 \begin{itemize}
 \item
  ${\bf \Delta_{12}\simeq0}$. Eq.~(\ref{disap}) reduces then to -for instance for the case of  $\bar{\nu}_e$ disappearance and  $\nu_\mu$ disappearance-
\bea
\nonumber
\hat{P}_{\bar\nu_e \bar\nu_e}&\simeq&
(|N_{e 1}|^2 + |N_{e 2}|^2)^2 + |N_{e 3}|^4 \\
&+& 2(|N_{e1}|^2+|N_{e2}|^2)|N_{e3}|^2 \cos \Delta_{23}\,,
\label{chooz}
\eea
\bea
\nonumber
\hat{P}_{\nu_\mu \nu_\mu}&\simeq&
(|N_{\mu 1}|^2 + |N_{\mu 2}|^2)^2 + |N_{\mu 3}|^4 \\
&+& 2(|N_{\mu1}|^2+|N_{\mu2}|^2)|N_{\mu3}|^2 \cos \Delta_{23}\,,
\label{atmo}
\eea
respectively. Relevant experiments in this class include
 CHOOZ~\cite{CHOOZ}, a reactor experiment sensitive to  $\bar\nu_e$ disappearance, as well as the $\nu_\mu$ disappearance atmospheric \cite{SK} and accelerator experiments such as K2K~\cite{K2K} or MINOS~\cite{Minos}. Eqs.~(\ref{chooz}) and (\ref{atmo}) indicate that this type of vacuum experiments cannot disentangle by themselves the element $|N_{\alpha 1}|$ from $|N_{\alpha 2}|$, as they appear in the combinations
$|N_{e 1}|^2 + |N_{e 2}|^2$ and $|N_{\mu 1}|^2 + |N_{\mu 2}|^2$, respectively.
In addition, the equations show as well the presence of a degeneracy between those combinations versus  $|N_{e 3}|$ and $|N_{\mu 3}|$, respectively.
  \item   {\bf ${\bf \Delta_{12}\ne0}$ with ${\bf \Delta_{23}\gg1}$.}  The latter -atmospheric- oscillation frequency is averaged out resulting in -for instance for $\bar\nu_e$ disappearance-
  \bea
\hat{P}_{\bar\nu_e \bar\nu_e}&\simeq&
|N_{e 1}|^4 + |N_{e 2}|^4 + |N_{e 3}|^4 + 2|N_{e1}|^2|N_{e2}|^2 \cos \Delta_{12}.
\label{kland}
\eea 
KamLAND~\cite{Kamland} is a reactor experiment with a longer baseline than CHOOZ and falling into this category. Notice that the dependence on $|N_{e 1}|$ and $|N_{e 2}|$ in Eq.~(\ref{kland})
 differs from that in Eq.~(\ref{chooz}), suggesting that the combination of both type of experiments may help to tell those elements apart, as it will be shown later on.
 \item {\bf ${\bf \Delta_{12}\simeq0}$ and ${\bf \Delta_{23}\simeq0}$}. The appearance and disappearance probabilities correspond  then to 
 a simple formula (see
  Eq.~(\ref{zerodist})):
 \bea
\hat{P}_{\nu_\alpha  \nu_\beta}&\simeq& |(NN^\dagger)_{\beta\alpha}|^2 \,.
\label{app-near}
\eea
KARMEN~\cite{KARMEN} and NOMAD~\cite{NOMAD} are appearance experiments in this class, well described by Eq.~(\ref{app-near}); the same holds for the data on $\nu_\mu$ disappearance at the near detector in MINOS and on $\bar\nu_e$ disappearance at BUGEY~\cite{BUGEY}.
 \end{itemize}
  
  \subsubsection*{ Oscillations in matter}
  
  A very important experiment in this class is SNO.
  In the unitary treatment, the
$\nu_e$ produced at the core of the sun are approximately eigenstates of the total Hamiltonian, since the interaction Hamiltonian dominates the evolution in this region of the sun. Denoting the total eigenstates in matter by $| \tilde{\nu}_i \rangle$, a $\nu_e$ in the center of the sun is approximately~\cite{sunnu2}:
\bea
|\nu_e \rangle &\simeq&
\sqrt{0.1} | \tilde{\nu}_1 \rangle+ \sqrt{0.9} | \tilde{\nu}_2 \rangle\,,
\label{snostate}
\eea
within a $2\%$ accuracy.
The state then evolves adiabatically so that, when leaving  the sun, the $| \tilde{\nu}_i \rangle$ states can be replaced by the vacuum eigenstates $| \nu_i \rangle$, leading to
\bea
\hat{P}_{\nu_e\nu_e}&\simeq&
0.1|N_{e 1}|^2 + 0.9|N_{e 2}|^2\,,
\label{sno}
\eea
which allows a clean measurement of $|N_{e 2}|^2$.
  
 Within the MUV scheme, {\it a priori} the analysis varies. This was illustrated in 
  Eq.~(\ref{Eflav-int}) for two-family oscillations in matter, which exhibits exotic non-diagonal terms and where the neutral currents may play {\it a priori} a significant role.  Nevertheless, we will see below that the absence of oscillation signals at near detectors  constrain deviations from unitarity, for all $(NN^\dagger)$ elements but  $(NN^\dagger)_{\tau\tau}$, to be smaller than ${\cal O}(10^{-1})$. In fact, it will turn out that all  bounds on $(NN^\dagger)$, including $(NN^\dagger)_{\tau\tau}$, are improved also from weak decays and the values of the off-diagonal elements constrained to be smaller than $5\%$, as it will be shown in the next Section. In consequence, for the level of precision aimed at in this work, it is unnecessary to perform the complete MUV analysis of SNO data and Eq.~(\ref{sno}) keeps being an appropriate approximation.
   This determination of $|N_{e2}|$ will be a major input in resolving the MUV degeneracy between $|N_{e1}|$ and $|N_{e2}|$.
    
 \vspace{0.7cm}
   In all numerical analysis below, the values of $\Delta m_{12}$ and  $\Delta m_{23}$,
   resulting from our fits in the MUV scheme, will not be shown: they coincide with those obtained in the unitary treatment, as expected from the fact that the oscillation frequencies are not modified in the MUV scheme, unlike the amplitudes.

\subsection{Constraints on the $\boldsymbol{e}$-row}
\label{fit-e}
\begin{figure}[t]
\vspace{-0.5cm}
\begin{center}
\begin{tabular}{cc}
\hspace{-0.55cm} \includegraphics[width=7.5cm]{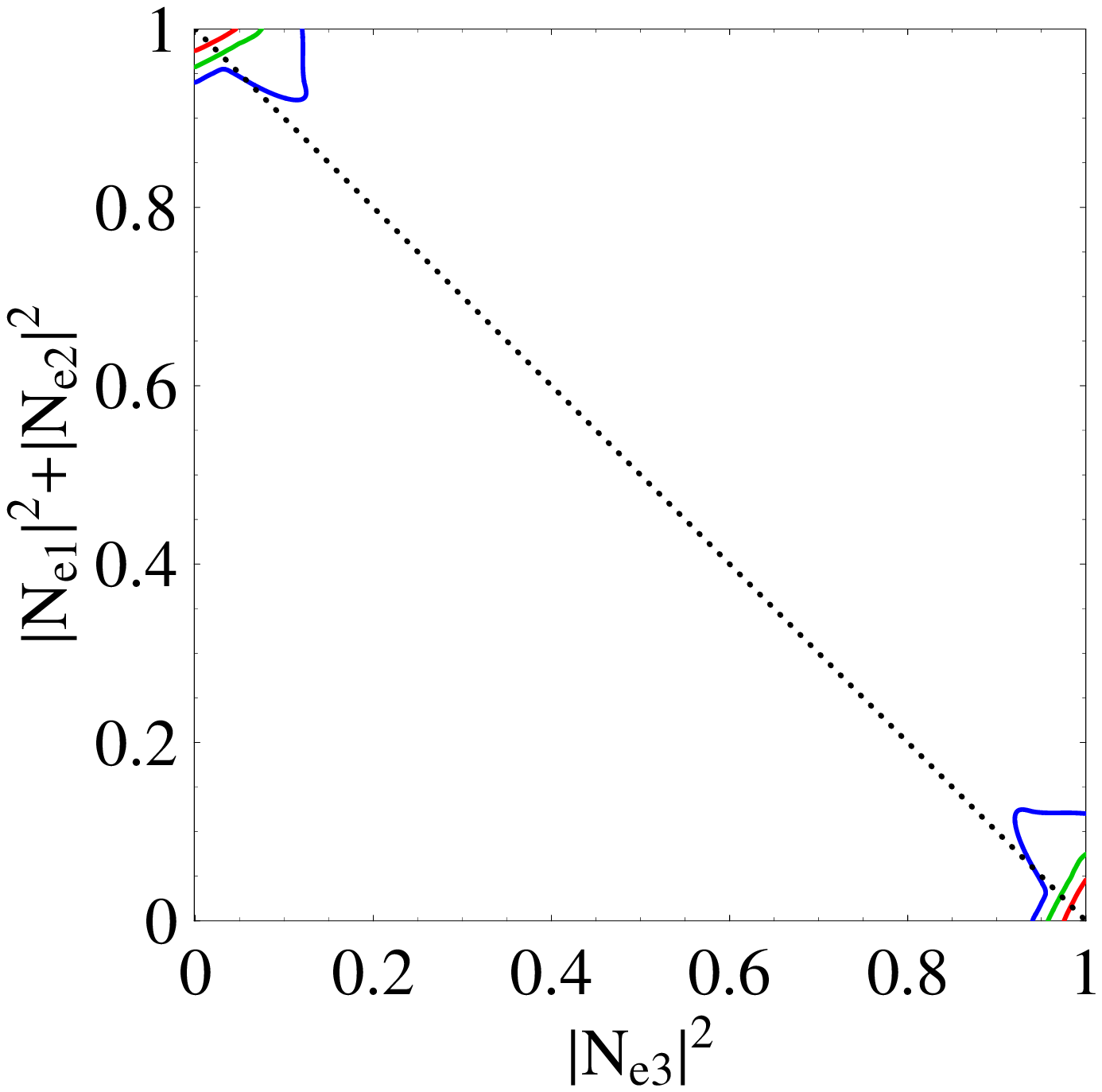} &
		 \includegraphics[width=7.5cm]{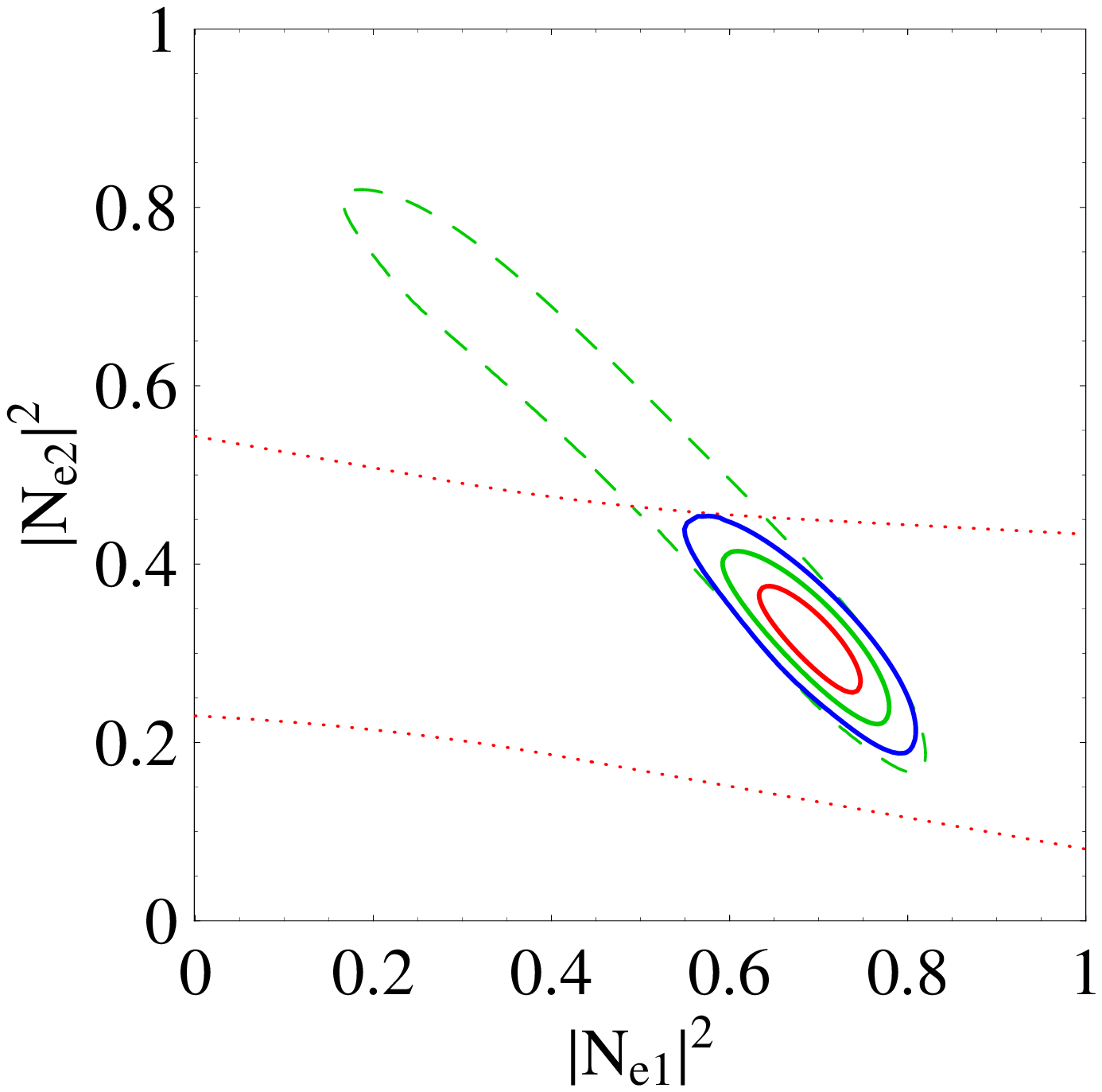}
\end{tabular}
\caption{\it Left: $1$, $2$ and $3 \sigma$ contours for CHOOZ and K2K data (solid
lines) with unitarity condition (dotted line).  Right: $3 \sigma$
contours for CHOOZ, K2K and KamLAND data (dashed line), SNO data (dotted
line) and their combination at $1$, $2$ and $3\sigma$ (solid lines).}
\label{fig:erow}
\end{center}
\end{figure}

In Fig.~\ref{fig:erow} (left) we present the 1, 2 and $3\sigma$ contours of a
three-family fit to CHOOZ data, combined with the information on
$\Delta m^2_{23}$ resulting from an analysis of K2K data. The dotted line
represents the unitarity condition $(NN^\dagger)_{ee} =1$. 
 
Since CHOOZ data are compatible with the
no-oscillation hypothesis, the fit shows allowed regions in which the first line in Eq.~(\ref{chooz}) is close to one, while
the second
-oscillatory- term vanishes. That is, either $|N_{e 1}|^2 + |N_{e
2}|^2 \simeq 1$ with $|N_{e 3}|^2 \simeq 0$, or $|N_{e 1}|^2 + |N_{e
2}|^2 \simeq 0$ with $|N_{e 3}|^2 \simeq 1$. The detection of the $L/E$ dependence in KamLAND selects the first combination, though, see Eq.~(\ref{kland}). The significant loss of sensitivity to $|N_{e3}|$ of the $3\sigma$ contour with respect to the $1$ and $2\sigma$ ones can be understood from the fact that CHOOZ loses its sensitivity for $\Delta m^2_{23} \simeq 0.001$, as can be seen in Fig.~55 of Ref.~\cite{CHOOZ}. Indeed K2K excludes such small values of $\Delta m^2_{23}$ at $1$ and $2\sigma$, but not at $3\sigma$, where the loss of sensitivity occurs. Notice that the $3\sigma$ contour intersects the
unitarity condition at $|N_{e 3}|^2\simeq 0.05$, which agrees with the
usual bounds for $|N_{e 3}|^2$, obtained under the assumption of unitarity.

KamLAND also helps to disentangle $|N_{e 1}|^2$ from $|N_{e 2}|^2$,  as illustrated
in Fig.~\ref{fig:erow} (right), in which the $3\sigma$ contour of a fit to
KamLAND data is presented (dashed line), combined with those from CHOOZ and K2K. Since CHOOZ constrained $|N_{e 1}|^2 + |N_{e
2}|^2$ to be close to  1,
 only a narrow strip near the diagonal is allowed. The region is still  large, though, due to the symmetry of
Eq.~(\ref{kland})  under the interchange of $|N_{e 1}|^2$
with $|N_{e 2}|^2$.

This final degeneracy can be lifted with information from SNO. 
The SNO data on the ratio of the charged-current over neutral-current fluxes 
results in the rather horizontal $3\sigma$ strip (dotted
line) in Fig.~\ref{fig:erow} (right).  To determine this region,
the ratio of charged-current over neutral-current fluxes~\cite{SNO}
can be approximated by Eq.~(\ref{sno}). A $5\%$ variation has been allowed, to take into account the corrections stemming from Eqs.~(\ref{eventos2amp}) and (\ref{Eflav-int}).
Furthermore, we have verified that even a $10\%$ variation in those coefficients would not change significantly the results of the fit.

The combined fit of CHOOZ, KamLAND, SNO and K2K data is
depicted at 1, 2 and $3 \sigma$ by the solid contours in Fig.~\ref{fig:erow} (right).  The Figure shows then that the combination of all this complementary
information  constrains all  elements of the $e$-row with a precision only slightly inferior to that  of the usual
unitary analysis.

\subsection{Constraints on the $\boldsymbol{\mu}$-row}
\label{fit-mu}
\begin{figure}[t]
\vspace{-0.5cm}
\begin{center}
\begin{tabular}{c}
\hspace{-0.55cm} \includegraphics[width=7.5cm]{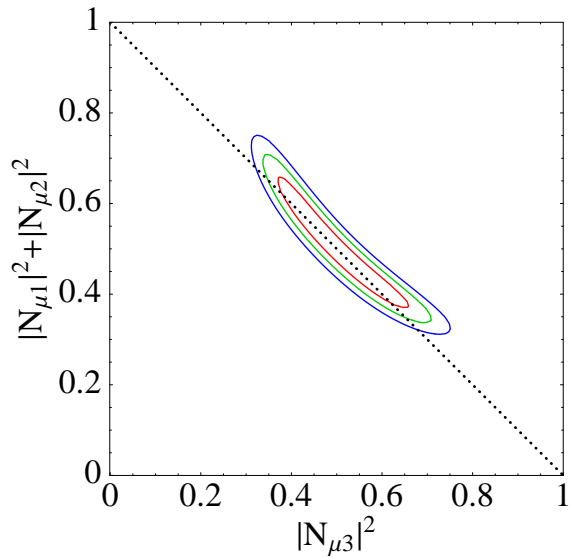}
\end{tabular}
\caption{\it $1$, $2$ and $3 \sigma$ contours for K2K and SK data (solid line) with
unitarity condition (dotted line).}
\label{fig:atmo}
\end{center}
\end{figure}
In Fig.~\ref{fig:atmo} we show the 1, 2 and $3\sigma$ contours (solid lines) of a
fit to K2K  data, combined with an estimation for SK. The latter resulted from translating the measured value of 
 $\sin^2(2\theta_{23})$ in Ref.~\cite{SK}
to matrix elements, using Eq.~(\ref{atmo}). The dotted line
represents the unitarity condition, $(NN^\dagger)_{\mu\mu}=1$. Without additional information at
different $L/E$, $|N_{\mu 1}|^2$ and $|N_{\mu 2}|^2$ can not be disentangled and our  knowledge of the $\mu$-row is much worse than when
imposing unitarity.

\vspace{0.5cm}

Putting together all the information developed above from  different oscillation experiments, 
 the following allowed ranges are obtained -at $3\sigma$-, for the elements of the leptonic mixing matrix:
\bea
|N| = \left( \begin{array}{ccc}
0.75-0.89 & 0.45-0.66 & <0.34 \\
\big[\,(|N_{\mu1}|^2+|N_{\mu2}|^2)^{1/2}= & 0.57-0.86\,\big] & 0.57-0.86 \\
? & ? & ? \end{array}
\right).
\label{N_osc}
\eea

We observe that, without the assumption of unitarity, present oscillation signals can only determine half of the matrix elements. The elements
of the first row have all been determined or constrained ($|N_{e2}|^2$
mainly by SNO, $|N_{e3}|^2$ mainly by CHOOZ and $|N_{e1}|^2$ by
KamLAND combined with the others). In contrast, for the second row,
present atmospheric and accelerator experiments are unable to
discriminate between $|N_{\mu
1}|^2$ and $|N_{\mu 2}|^2$.
Finally, no direct information can be obtained on the
$\tau$-row in the absence of $\nu_\tau$ oscillations signals.

\subsection{Constraints on non-unitarity from near detectors}
 NOMAD, KARMEN, BUGEY and the near detector at MINOS provide constraints on the elements of $NN^\dagger$
as it follows from Eq.~(\ref{app-near}). We obtain, at the $90\%$CL,
\begin{eqnarray}
|N N^\dagger | \approx
\begin{pmatrix}
 1.00\pm  0.04   & < 0.05  &   < 0.09\\
 < 0.05   &  1.00 \pm  0.05 &  < 0.013  \\
 < 0.09   &  < 0.013  &  ?
\end{pmatrix} \, .
\label{nndagzero}
\end{eqnarray}

\vspace{0.7cm}

With this information, $|N_{\mu 1}|$ and $|N_{\mu 2}|$  in Eq.~(\ref{N_osc}) can now be disentangled. All in all, the mixing matrix resulting from analyzing oscillation data within the MUV scheme is given by
\bea
|N| = \left( \begin{array}{ccc}
0.75-0.89 & 0.45-0.66 & <0.27 \\
0.00-0.69 & 0.22-0.81 & 0.57-0.85 \\
? & ? & ?
\end{array}
\right)\, .
\label{N_zero}
\eea
Notice that, even adding the constraints obtained at near detectors, not all matrix elements can be determined from oscillation data.

\section{Constraints on non-unitarity from electroweak\\ decays}
\label{decays}

 Neutrino oscillations are evidence of a non-trivial leptonic mixing, allowing to determine the individual elements of the mixing matrix from its data, as done in the previous Section. In contrast, leptonic and semileptonic decay data are not appropriate for this task.   This is because,  contrary to the quark sector, where the different quark mass eigenstates can be tagged, neutrino eigenstates are not detected separately.
  The experimentally measured rates correspond then to sums over all possible mass eigenstates, resulting only in sums of {\it products} of matrix elements.

  Leptonic and semileptonic decays may be sensitive to leptonic non-unitarity, though,
  due to the ``zero-distance'' effect, which in the flavour basis is encoded by its non-orthogonality, Eq.~(\ref{noortoflav}). The combinations $(NN^\dagger)_{\alpha\beta}$ can be extracted from them, as suggested by Eqs.~(\ref{xcc1})-(\ref{xnc}).
  With that aim, $W$, $Z$, $\pi$ and lepton decays are analyzed in this chapter, in the MUV scheme. The results will further constrain the mixing matrix obtained from neutrino oscillation processes.

\subsection{ $\boldsymbol{W}$ decays}

\label{W-decays}
With a non-unitary leptonic mixing matrix $N$, the decay widths for $W$
into charged leptons and neutrinos are given -as in Eq.~(\ref{xcc1})- by
\begin{eqnarray}
\label{Eq:Wdecay1}
\Gamma (W \rightarrow \ell_\alpha \nu_\alpha ) =
\sum_i\Gamma (W \rightarrow \ell_\alpha \nu_i ) =
\frac{G_F M_W^3}{6 \sqrt{2} \pi} (N N^\dagger)_{\alpha\alpha} \, .
\end{eqnarray}
 $G_F$  has been related to the Fermi constant $G^M_F$, measured from the decay $\mu \rightarrow \nu_\mu e
\bar{\nu}_e$,  by Eq.~(\ref{Eq:GFM}), allowing to extract now from Eq.~(\ref{Eq:Wdecay1}) the following combinations:
\begin{eqnarray}
\frac{(NN^\dagger)_{\alpha\alpha}}{\sqrt{(NN^\dagger)_{ee}(NN^\dagger)_{\mu\mu}}}
= \frac{\Gamma (W \rightarrow \ell_\alpha \nu_\alpha ) \, 6 \sqrt{2} \pi}{G^M_F M_W^3}
\equiv f_\alpha \, .
\end{eqnarray}
Using the results for the $W$ decay widths and mass from Ref.~\cite{PDG}, as well as $G^M_F = (1.16637 \pm 0.00001) \cdot 10^{-5}$, the parameters
$f_\alpha$ are
\begin{eqnarray}
\nonumber
f_e &=& 1.000\pm  0.024\, ,\\
\nonumber
f_\mu &=& 0.986 \pm  0.028\, ,\\
f_\tau &=& 1.002 \pm 0.032 \, .
\end{eqnarray}
\subsection{ Invisible $\boldsymbol{Z}$ decay}
\label{Z-decays}
Further constraints stem from the invisible $Z$-decay width, which,
for non-unitary leptonic mixing $N$, is given by (see Eq.~(\ref{xnc}))
\begin{eqnarray}
\label{iag}
\Gamma (Z \rightarrow \mbox{invisible} ) =  \sum_{i,j}\Gamma (Z \rightarrow \bar{\nu}_i \nu_j ) = \frac{G_F M_Z^3}{12 \sqrt{2} \pi}\ (1 + \rho_t)
\,\sum_{i,j} |(N^\dagger N)_{ij}|^2\,,
\end{eqnarray}
where $\rho_t \approx 0.008$~\cite{PDG} takes into account radiative corrections mainly steming from loops with the top quark. As the dominant radiative corrections do not involve neutrinos, the dependence on the mixing matrix in Eq.~(\ref{iag}) appears as a global factor to an excellent approximation. 
Using Eq.~(\ref{Eq:GFM}), the equality $\sum_{i,j} |(N^\dagger
N)_{ij}|^2=\sum_{\alpha,\beta} |(N N^\dagger)_{\alpha\beta}|^2$ and
the data provided in Ref.~\cite{PDG}, the following constraint is
obtained
\begin{eqnarray}
\frac{\sum_{\alpha,\beta} |(N N^\dagger)_{\alpha\beta}|^2}{\sqrt{(NN^\dagger)_{ee}(NN^\dagger)_{\mu\mu}}}
= \frac{12 \sqrt{2} \pi \,\Gamma (Z \rightarrow \mbox{invisible} )}{G_F^M M_Z^3(1+\rho_t)}
= 2.984 \pm 0.009 \,.
\end{eqnarray}
As it is well known, this number should correspond to the number of active neutrinos at LEP. Its $2\sigma$ departure from the value of $3$ is not (yet) significant enough 
to be interpreted as a signal of new physics~\footnote{We thank B. Kayser and J. Kersten for reminding us of this departure from the SM prediction as a possible signal of non-unitarity and  of the importance of  radiative corrections in this process, respectively.}.
\subsection{ Universality tests}
\label{univ-tests}
In addition, ratios of lepton, $W$ and $\pi$ decays, used often in
order to test universality~\cite{PDG,universality}, can be interpreted as tests of lepton
mixing unitarity.  They result in
constraints for the diagonal elements of $NN^\dagger$, as suggested by
Eqs.~(\ref{xcc1})-(\ref{xnc}) and resumed in Table I.

\begin{table}[ht]
\centering 
\begin{eqnarray*} 
\begin{array}{|c|c|c|} 
\hline 
\displaystyle \mbox{Constraints on}  \vphantom{\frac{a}{a}}
&
\mbox{Process}
&
\mbox{Bound}
\\
\hline\hline
\displaystyle\frac{(N N^\dagger)_{\mu\mu}}{(N N^\dagger)_{ee}}  
&
\displaystyle\frac{\Gamma (\tau \rightarrow \nu_\tau \mu \bar{\nu}_\mu )}{
      \Gamma (\tau \rightarrow \nu_\tau e \bar{\nu}_e )}
&
0.9999 \pm 0.0020
\\
\hline 
\displaystyle\frac{(N N^\dagger)_{\mu\mu}}{(N N^\dagger)_{ee}}  
&
\displaystyle\frac{\Gamma (\pi \rightarrow \mu \bar{\nu}_\mu )}{
      \Gamma (\pi \rightarrow e \bar{\nu}_e )}
&
1.0017 \pm 0.0015
\\
\hline
\displaystyle\frac{(N N^\dagger)_{\mu\mu}}{(N N^\dagger)_{ee}}
&
\displaystyle\frac{\Gamma (W \rightarrow \mu \bar{\nu}_\mu )}{
      \Gamma (W \rightarrow e \bar{\nu}_e )}
&
0.997 \pm 0.010
\\
\hline
\displaystyle\frac{(N N^\dagger)_{\tau\tau}}{(N N^\dagger)_{\mu\mu}}
&
\displaystyle\frac{\Gamma (\tau \rightarrow \nu_\tau  e \bar{\nu}_e)}{
      \Gamma (\mu \rightarrow \nu_\mu e \bar{\nu}_e )}
&
1.0004 \pm 0.0023     
\\
\hline
\displaystyle\frac{(N N^\dagger)_{\tau\tau}}{(N N^\dagger)_{\mu\mu}}  
&
\displaystyle\frac{\Gamma (\tau \rightarrow \nu_\tau \pi)}{
      \Gamma (\pi \rightarrow \mu \bar{\nu}_\mu )}
&
0.9999 \pm 0.0036
\\
\hline
\displaystyle\frac{(N N^\dagger)_{\tau\tau}}{(N N^\dagger)_{ee}}  
&
\displaystyle\frac{\Gamma (\tau \rightarrow \nu_\tau \mu \bar{\nu}_\mu )}{
      \Gamma (\mu \rightarrow \nu_\mu e \bar{\nu}_e )}
&
1.0002 \pm 0.0022
\\
\hline
\displaystyle\frac{(N N^\dagger)_{\tau\tau}}{(N N^\dagger)_{ee}}  
&
\displaystyle\frac{\Gamma (W \rightarrow \tau \bar{\nu}_\tau )}{
      \Gamma (W \rightarrow e \bar{\nu}_e )}
&
1.034 \pm 0.014
\\
\hline
\end{array} 
\end{eqnarray*} 
\caption{
\label{tab:UnivTests} \it Constraints on $(N N^\dagger)_{\alpha\alpha}$ from a selection of processes.} 
\end{table}

The processes investigated so far constrained the diagonal elements of the product $N N^\dagger$. Limit values for its off-diagonal elements can be obtained instead from rare decays of  charged leptons, as we show next.

\subsection{Rare charged lepton decays}
\label{rare}
\begin{figure}[t]
\vspace{-0.5cm}
\begin{center}
\begin{tabular}{c}
\hspace{-0.55cm} \includegraphics[width=6.5cm]{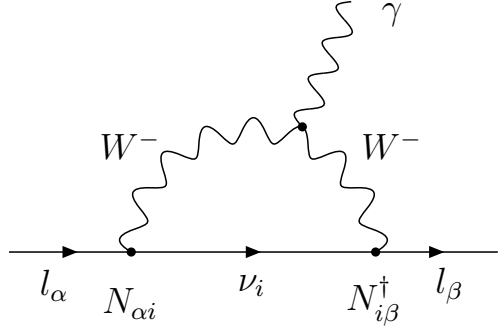}
\end{tabular}
\caption{\it One-loop diagram contributing to rare lepton decays.}
\label{muegamma}
\end{center}
\end{figure}
The leptonic process $\ell_\alpha \rightarrow \ell_\beta \gamma$ only occurs at one loop, as illustrated in Fig.~\ref{muegamma}. As the photon is on-shell, there are no divergent contributions to the diagram. The
1-loop branching ratio in the MUV scheme is given by the same expression~\cite{meg_sm} as in  the unitary case, substituting $U_{PMNS}$ by $N$,
\begin{eqnarray}
%Br (\ell_\alpha \rightarrow \ell_\beta \gamma) = \
\frac{\Gamma (\ell_\alpha \rightarrow \ell_\beta \gamma)}{\Gamma (\ell_\alpha \rightarrow \nu_\alpha \ell_\beta \overline{\nu}_\beta)} = \frac{3 \alpha}{32 \pi}\,
\frac{ | \sum_k N_{\alpha k} N^\dagger_{k\beta} F (x_k) |^2 }{
(NN^\dagger)_{\alpha\alpha} (NN^\dagger)_{\beta\beta}}  ,
\end{eqnarray}
where $x_k \equiv m_k^2 / M_W^2$ with $m_k$ being the masses of
the light neutrinos and
\begin{eqnarray}
F(x)\equiv \frac{10 - 43 x + 78 x^2 - 49 x^3 + 4 x^4 + 18 x^3 \ln x}{3 (x-1)^4}\, .
\end{eqnarray}
 Would $N$ be unitary, the $x$-independent term would vanish exactly through the GIM mechanism~\cite{GIM}, for
$\alpha \not= \beta$. With $N$ non-unitary and $(N
N^\dagger)_{\alpha \beta} \not= \delta_{\alpha\beta}$, that term remains and can be the leading contribution to the branching ratio. With the -very accurate- approximation $F(x)\approx 10/3$,
it follows that
\begin{eqnarray}
%Br (\ell_\alpha \rightarrow \ell_\beta \gamma)
\frac{\Gamma (\ell_\alpha \rightarrow \ell_\beta \gamma)}{\Gamma (\ell_\alpha \rightarrow \nu_\alpha \ell_\beta \overline{\nu}_\beta)}= \frac{100 \alpha}{96 \pi} \frac{| (N N^\dagger)_{\alpha\beta}|^2}{(N N^\dagger)_{\alpha\alpha}
(N N^\dagger)_{\beta\beta}} \, ,
\end{eqnarray}
leading to the constraint
\begin{eqnarray}
\frac{| (N N^\dagger)_{\alpha\beta}|^2}{(N N^\dagger)_{\alpha\alpha}
(N N^\dagger)_{\beta\beta}} = \frac{\Gamma (\ell_\alpha \rightarrow \ell_\beta \gamma)}{\Gamma (\ell_\alpha \rightarrow \nu_\alpha \ell_\beta \overline{\nu}_\beta)} \frac{96 \pi}{100 \alpha} \, .
\end{eqnarray}
Strong constraints can now be obtained for the off-diagonal elements of $(NN^\dagger)$, using   the present experimental
bounds \cite{muegamma}-\cite{taumugamma}
\begin{eqnarray}
Br (\mu  \rightarrow e \gamma) &<& 1.2 \cdot 10^{-11}
%\cdot (\Gamma_{\mathrm{full}}^\tau \,/\, \Gamma (\tau \rightarrow \nu_\tau \mu \overline{\nu}_\mu))
\; ,\\
Br (\tau \rightarrow e \gamma) &<& 1.1 \cdot 10^{-7}
%\cdot (\Gamma_{\mathrm{full}}^\tau \,/\, \Gamma (\tau \rightarrow \nu_\tau e \overline{\nu}_e))
\; ,\\
Br (\tau \rightarrow \mu \gamma) &<& 4.5 \cdot 10^{-8}
%\cdot (\Gamma_{\mathrm{full}}^\tau \,/\, \Gamma (\mu \rightarrow \nu_\mu e \overline{\nu}_e))
\; ,
\end{eqnarray}
together with the experimental values $Br (\tau \rightarrow \nu_\tau \mu
\overline{\nu}_\mu) = 0.1736 \pm
0.0006$, $Br (\tau \rightarrow \nu_\tau e \overline{\nu}_e) = 0.1784 \pm 0.0006$ and $Br (\mu
\rightarrow \nu_\mu e \overline{\nu}_e) \approx 100$\%~\cite{PDG}.

The strong experimental bound on $\mu \rightarrow e \gamma$ results in\footnote{This
strong bound also rules out the possibility of explaining the LSND
anomaly with the ``zero-distance'' effect, at least in our minimal
scheme.} 
$(NN^\dagger)_{e\mu(\mu e)}< 10^{-4}$, while the other off-diagonal
elements are constrained to be less than a few percent.

Finally, other lepton-flavour violating decays like $\ell_i \rightarrow
3\ell $, as well as $\ell_i \rightarrow \ell_j$ conversion in nuclei,
impose additional constraints close to those above. They may become
increasingly relevant, depending on the experimental sensitivities
attained in the future, as it will be discussed in Sec.~\ref{future}.

\subsection{Summary of constraints on non-unitarity from decays}
\label{summary}
All in all, a global fit to the constraints listed in this Section proves that the $NN^\dagger$ elements
 agree with those expected in the unitary case,
within a precision better than a few percent, at the 90\% CL:
\begin{eqnarray}
|N N^\dagger | \approx
\begin{pmatrix}
 0.994\pm  0.005   & < 7.0 \cdot 10^{-5}  &   < 1.6 \cdot 10^{-2}\\
 < 7.0 \cdot 10^{-5}   &  0.995 \pm  0.005 &  < 1.0 \cdot 10^{-2}  \\
 < 1.6 \cdot 10^{-2}   &  < 1.0 \cdot 10^{-2}  &  0.995 \pm 0.005
\end{pmatrix} \, .
\label{nndag}
\end{eqnarray}
Notice that there is a $2\sigma$ departure from the value of $1$ in the diagonal elements of this matrix, stemming from the similar discrepancy in the invisible width of the $Z$ boson, not (yet) significant enough 
to be interpreted as a signal of new physics.

  In contrast, there is no direct information from decays on the product $N^\dagger N$, except
that resulting from the invisible decay width of the $Z$ boson. 
We can infer strong limits on the $N^\dagger N$ elements, though, from
those in Eq.~(\ref{nndag}).  Parametrize the matrix $N$
as $N \equiv H \,V$, where $V$ is a unitary matrix and $H$ Hermitian,
\begin{eqnarray}
N N^\dagger = H^2 \equiv 1 + \varepsilon
\end{eqnarray}
with $\varepsilon=\varepsilon^\dagger$
and
\begin{eqnarray}
N^\dagger N = V^\dagger
H^2 V = 1 + V^\dagger \varepsilon V \equiv 1 + \varepsilon'\;.
\end{eqnarray}
 $\varepsilon$ ($\varepsilon'$) parameterizes
the allowed deviation of $N N^\dagger$ ($N^\dagger N$) from the unit matrix. It follows that
\begin{eqnarray}
|\varepsilon'_{i j}|^2 \le 
\sum_{i j} |\varepsilon'_{i j}|^2 = 
\sum_{\alpha\beta}  | \varepsilon_{\alpha\beta} |^2 \,,
\end{eqnarray}
where the unitarity of $V$ has been used, resulting in the constraint
\begin{eqnarray}|\varepsilon'_{i j}| \le (\sum_{\alpha \beta} | \varepsilon_{\alpha \beta} |^2)^{1/2}
= 0.032\;.
\end{eqnarray}
$N^\dagger N$ is thus constrained as follows:
\begin{eqnarray}
|N^\dagger N | \approx
\begin{pmatrix}
 1.00\pm  0.032   & < \,\,0.032 \,\,\, &   < 0.032\\
 < 0.032   &  1.00 \pm  0.032 &  < 0.032  \\
 < 0.032   &  < 0.032  &  1.00 \pm 0.032
\end{pmatrix} \, .
\label{ndagn}
\end{eqnarray}

 The results in Eqs.~(\ref{nndag}) and (\ref{ndagn}) prove that,
within the MUV scheme, unitarity in the lepton sector is
experimentally confirmed from data on weak decays with a precision
better than $5\%$, and does {\em not} need to be imposed as an
assumption, within that accuracy.  This means as well that the
leptonic unitarity triangles~\cite{Utriangles,Farzan:2002ct} and
normalization conditions -corresponding to the elements of $N
N^\dagger$ and $N^\dagger N$- are experimentally checked with a
precision of a few $\%$ (or much higher, as for instance for the $\mu$
- $e$ triangle).

\section{The mixing matrix}
\label{totalmixmatrix}
 The elements of the mixing matrix obtained from the analysis of neutrino oscillation experiments, Eq.~(\ref{N_zero}), can now be  combined  with the unitarity constraints obtained from weak decays in Eqs.~(\ref{nndag}) and (\ref{ndagn}).  The resulting mixing matrix in the MUV scheme is
\bea
|N| = \left( \begin{array}{ccc}
0.75-0.89 & 0.45-0.65 & <0.20 \\
0.19-0.55 & 0.42-0.74 & 0.57-0.82 \\
0.13-0.56 & 0.36-0.75 & 0.54-0.82
\end{array}
\right)\,.
\label{N_dec}
\eea
All the elements are now significantly constrained to be rather close to those stemming from the usual unitary analysis \cite{concha},
\bea
|U| = \left( \begin{array}{ccc}
0.79-0.88 & 0.47-0.61 & <0.20 \\
0.19-0.52 & 0.42-0.73 & 0.58-0.82 \\
0.20-0.53 & 0.44-0.74 & 0.56-0.81 \end{array}
\right)\, .
\label{U}
\eea

The constraints resulting for the $N_{e1}$ and $N_{e2}$ elements are
somewhat looser than their partners in the unitary analysis. This is due to the large uncertainties allowed for the values of the coefficients in
Eq.~(\ref{snostate}), together with the fact that, among all data available from solar experiments, we have only included in our analysis the SNO ratio of charged to neutral current events. Notice also that
the elements of the $\tau$-row are significantly less bounded than in
the unitary analysis, their values being inferred only indirectly.

%%%%%%   section 6   %%%%%%%%%%%%%%%%%%%%%%%%%%%%%%%%%%%%%%%%%%%%%%%
\section{Future experiments}
\label{future}

\subsubsection*{Matrix elements}

In order to measure independently $|N_{\mu 1}|^2$ and $|N_{\mu 2}|^2$
without relying on indirect information, a $\nu_{\mu}$ disappearance
experiment sensitive to $\Delta m^2_{12}$ (with the oscillations
driven by $\Delta m^2_{23}$ averaged out), as the one proposed in
Ref.~\cite{Farzan:2002ct}, would be needed, as suggested by
Eq.~(\ref{kland}) replacing $e$ by $\mu$. This experiment is quite
challenging, requiring an intense $\nu_\mu$ beam of low energy
($\simeq 500$ MeV) and a very long baseline ($\simeq 2000$ km).

Future facilities under discussion, searching for CP-violation in
appearance channels, include Super-Beams~\cite{T2K},
$\beta$-Beams~\cite{Zucchelli:sa} and Neutrino
Factories~\cite{nufact}.  The two latter ones may study $\nu_e \to
\nu_\mu$ and $\bar \nu_e \to \bar \nu_\mu$ transitions~\cite{golden}
with great precision, while Super-Beams may explore their T-conjugate
channels. Unlike disappearance experiments, measurements at these
facilities will thus be sensitive to the phases of the matrix
elements, which at present remain completely unknown. Neutrino Factory
beams would also be energetic enough for the $\nu_e \to \nu_\tau$ and
$\nu_\mu \to \nu_\tau$ oscillation channels to be
accessible~\cite{burguet,Donini:2002rm}. The $\tau$-row could thus be
tested directly and without relying on indirect decay information.

\subsubsection*{$\boldsymbol{(NN^\dagger)_{\mu e}}$}
As regards unitarity bounds, the constraints on non-unitarity from
decays are also likely to improve. If no positive signal is found for
$\mu \to e \gamma$, the bound on its branching ratio is expected to
reach $2\cdot10^{-14}$ in the near future~\cite{muegammafuture}, which
can be translated into a unitarity constraint $(NN^\dagger)_{\mu
e}<2.9\cdot10^{-6}$.  At a Neutrino Factory, the branching ratio for
$\mu \to e \gamma$ could be further constrained to
$<10^{-15}$~\cite{sensnufact}, which would result in
$(NN^\dagger)_{\mu e}<6.4\cdot10^{-7}$, at the $90\%$CL.

Important improvements are also expected regarding the bounds for
$\mu$ to $e$ conversion in nuclei. This process is more suppressed
than $\mu \to e \gamma$, though, due to the extra electromagnetic
coupling. In a Neutrino Factory, sensitivities down to $10^{-18}$
could be achieved~\cite{sensnufact} which, translated to
$(NN^\dagger)_{\mu e}\simeq 3.2\cdot10^{-7}$, are only a factor two
stronger than the bound expected from $\mu \to e \gamma$. Similar
ultimate sensitivities are being discussed as regards the PRISM/PRIME
project~\cite{PRIME}.

\subsubsection*{$\boldsymbol{(NN^\dagger)_{e\tau}}$ and $\boldsymbol{(NN^\dagger)_{\mu \tau}}$}
On the other hand, the bounds on rare $\tau$ decays are not likely to
improve much without a dedicated facility, since Babar and Belle are
now limited by the background and an increase in statistics would not
improve the relevant measurements \cite{raretau}.

In contrast, the possibility of detecting $\nu_\tau$ at a near
detector of a Neutrino Factory would allow to improve the bounds on
$(NN^\dagger)_{e\tau}$ and $(NN^\dagger)_{\mu\tau}$. We have
considered an OPERA-like detector, located at a $100$m baseline from a
Neutrino Factory beam\footnote{This is only an example of the potential of detecting 
$\nu_\tau$ near the Neutrino Factory beam. A detailed study of whether the 
performance of an OPERA-like detector can be extrapolated to the neutrino luminosities 
so close to the source would be required, though.}, with a total mass of 4kt and the efficiencies
and backgrounds considered in Ref.~\cite{Donini:2002rm}. Assuming a
conservative $5 \%$ systematic error, the present bounds could be
improved to $(NN^\dagger)_{e\tau}<2.9\cdot 10^{-3}$ and
$(NN^\dagger)_{\mu\tau}<2.6\cdot10^{-3}$, at the $90\%$CL.

\section{Summary and conclusions}
\label{conclu}

The flavour mixing matrix present in leptonic weak currents may be
generically non-unitary, as a result of new physics, as for instance
that responsible for neutrino masses.  It is important to know up to
which point the values of the matrix elements are allowed by data to
differ from those obtained in the usual unitary analysis, as putative
windows of new physics.
 
Without attaching ourselves to any particular model, we have studied
a minimal scheme of unitarity violation -MUV-, considering only three
light neutrino species and with the usual unitary matrix $U_{PMNS}$
replaced by the most general non-unitary one.
   
We have first clarified the formalism for studying in this scheme
neutrino oscillations in vacuum and in matter. It results in
``zero-distance'' flavour-changing transitions, while neutrino
propagation in matter exhibits exotic couplings and an active role of
the neutral current couplings, unlike in the standard unitary
analysis.  The most relevant data from neutrino oscillation
experiments have then been used to determine, within the MUV scheme,
the elements of the mixing matrix: about half of its elements remain
unconstrained at this level.
    
Data from weak decays cannot determine the values of the elements of
the mixing matrix. They provide very stringent tests of unitarity,
though, which turns out to be satisfied at the level of a few
percent. The combination of nowadays' data from weak decays and
neutrino oscillation experiments results, within the MUV scheme, in a
set of absolute values for the elements of the mixing matrix, which is
very close to that from the unitary analysis. It is important to
pursue these searches with increasing precision, as it may well result
in a positive signal instead of  constraints, heralding new
physics.
  
Future facilities could establish the first signs of non-unitarity (or
new physics in general). We have discussed their potential to improve
unitarity constraints and determine all matrix elements without
assuming unitarity. While the unitarity bounds expected from
traditional $\mu\rightarrow e \gamma$ experiments will remain very
powerful, experiments at a Neutrino Factory can set also stringent
unitarity constraints, allow a direct determination of the elements in
the $\tau$ row of the mixing matrix and furthermore be sensitive to
the phases of the matrix elements.  More generally, the plethora of
flavour channels and precision measurements, offered by the several
types of future facilities under discussion, will be very significant
in the search for non-unitary effects as windows of new physics.
    
A final warning is pertinent here. The putative new physics may result
in modifications to other standard couplings than those considered
above, as well as in new exotic ones. Barring extreme fine-tunings in
some hypothetical model, our approach should provide, though, the
correct order of magnitude of the minimal unitarity violations allowed
by data.

\section*{Acknowledgments}

We are especially indebted to C. Pe\~na-Garay for illuminating
discussions on the analysis of oscillation data and for his
participation in the early stage of this work.  We also acknowledge
stimulating discussions with J.~Aguado, E.~Alba, W.~Buchm\"uller, 
A.~Donini, F.~Feruglio, C.~Gonz\'alez-Garc\'{\i}a, P.~Hern\'andez, B.~Kayser, J.~Kersten,
S.~Petcov, S.~Rigolin, L.~Rodrigo and M.~Salvatori. The work of
S.~Antusch was partially supported by the EU 6th Framework Program
MRTN-CT-2004-503369 ``The Quest for Unification: Theory Confronts
Experiment''. The work of C.~Biggio was partially supported by an INFN
postdoctoral fellowship. Furthermore, the authors received partial
support from CICYT through the project FPA2003-04597, as well as from
the Comunidad Aut\'onoma de Madrid through Proyecto HEPHACOS ;
P-ESP-00346. E.~Fern\'andez-Mart\'{\i}nez acknowledges the UAM for
financial support through a FPU fellowship. He acknowledges as well
the hospitality and financial support of the Institute for Advanced
Studies (Princeton), where part of his work was done.
%
 
%%%%%%   references   %%%%%%%%%%%%%%%%%%%%%%%%%%%%%%%%%%%%%%%%%%%%%%

\end{document}